\documentclass[pra,twocolumn,showpacs,preprintnumbers,amsmath,amssymb,superscriptaddress]{revtex4}

\usepackage{graphicx}
\usepackage{dcolumn}
\usepackage{bm}

\def \ed {\end{document}}
\def\Fbox#1{\vskip1ex\hbox to 8.5cm{\hfil\fboxsep0.3cm\fbox{%
  \parbox{8.0cm}{#1}}\hfil}\vskip1ex\noindent}  

\def\be{\begin{equation}}\def\ee{\end{equation}}
\def\bea{\begin{eqnarray}}\def\eea{\end{eqnarray}}
\def\bse{\begin{subequations}}\def\ese{\end{subequations}}
\newcommand{\BE}[1]{\begin{equation}\label{#1}}
\newcommand{\BEA}[1]{\begin{eqnarray}\label{#1}}
\newcommand{\BSE}[1]{\begin{subequations}\label{#1}}

\let \= \equiv \let\*\cdot \let\~\widetilde \let\^\widehat \let\-\overline

\def\<{\left\langle}    \def\>{\right\rangle}
\def\({\left(}          \def\){\right)}
 \def \[ {\left [} \def \] {\right ]}

\makeatletter
\AtBeginDocument{\@ifpackageloaded{natbib}{\ifNAT@numbers\if@filesw\immediate\write\@auxout{\string\global\string\NAT@numberstrue}\fi\fi}{}}
\makeatother
\begin{document}

\preprint{APS/123-QED}

\title{Spin turbulence with small spin magnitude in spin-1 spinor Bose-Einstein condensates} 

\author{Kazuya Fujimoto}
\affiliation{Department of Physics, Osaka City University, Sumiyoshi-ku, Osaka 558-8585, Japan}

\author{Makoto Tsubota}
\affiliation{Department of Physics, Osaka City University, Sumiyoshi-ku, Osaka 558-8585, Japan}
\affiliation{The OCU Advanced Research Institute for Natural Science and Technology (OCARINA), Osaka City University, Sumiyoshi-ku, Osaka 558-8585, Japan}

\date{\today}

\begin{abstract}
We theoretically and numerically study spin turbulence (ST) with small spin magnitude in spin-1 spinor Bose-Einstein condensates  by using the spin-1 spinor 
Gross-Pitaevskii (GP)equations. This kind of ST is realized in  two cases:
(i)  with antiferromagnetic  interaction and (ii) with ferromagnetic  interaction under a static magnetic field. 
The ST with small spin magnitude can exhibit two characteristic power laws in the spectrum of the spin-dependent interaction energy:  $-1$ and $-7/3$ power laws in the low- and high-wave-number regions, respectively. These power laws are derived from a Kolmogorov-type dimensional scaling analysis for the equations of motion of the spin vector and nematic tensor. 
To confirm these power laws, we perform a numerical calculation of the spin-1 spinor GP equations in a two-dimensional uniform system.
In  case (i), the $-7/3$ power law appears in the high-wave-number region, but the spectrum in the low-wave-number region deviates from the $-1$ power law. In contrast, both $-1$ and $-7/3$ power laws are found to clearly appear in  case  (ii). 
\end{abstract}

\pacs{03.75.Mn, 03.75.Kk}

\maketitle

\section{Introduction}  
Turbulent flow universally appears in various systems.
Generally, turbulence in a classical fluid [called classical turbulence (CT)] apparently seems to be very disordered and complex, but it is  known to exhibit some characteristic 
statistical laws. One of the most famous laws in CT is the Kolmogrov $-5/3$ power law, which is confirmed by the many numerical and 
experimental studies in fully developed isotropic turbulence \cite{Davidson,Frisch}. 

Quantum turbulence (QT), which is a turbulent state realized in a quantum fluid, has been studied in superfluid helium 
for a long time \cite{Hal}.
Recently, the study of QT in atomic Bose-Einstein condensates (BECs) has been an active area of research  \cite{KT07, Gou09, Henn09, White10, Boris12, Reeves12}. Some numerical studies of QT in atomic BECs have confirmed the Kolmogorov $-5/3$ power law in the spectrum of the incompressible kinetic energy \cite{KT07, Gou09, Reeves12}, which shows the analogy between CT and QT. 
However, there are some differences between CT and QT, an example of which is the velocity distribution \cite{White10}. 
In QT, a quantized vortex with  discrete circulation exists, making the velocity distribution different from the vortex in CT 
because the structure of the vortex core in quantum fluids is considerably different from that in classical ones. 
Thus, CT and QT exhibit both universal statistical laws independent of the details of the system and statistical laws characteristic of the system. 

In atomic BECs, there exist multicomponent BECs with  internal degrees of freedom that exhibit novel properties not seen in scalar BECs \cite{PR,KU,stamper}. 
Recently, hydrodynamics in binary BECs based on the Gross-Pitaevskii (GP) equations has been actively investigated, and various hydrodynamical instabilities, such as Rayleigh-Taylor, Kelvin-Helmholtz, and Richtmyer-Meshkov instabilities, have been  studied \cite{Sasaki09, Take10, Bezett10, Hamner11}. 
As another multicomponent system, spinor BECs are realized; these have  spin degrees of freedom and exhibit  behavior characteristic of spins \cite{KU,stamper}. 
As with binary BECs, some authors have studied the hydrodynamics in spinor BECs \cite{Lamacraft, Barnett, KK, YU12}, 
discussing spin dynamics such as the dynamical instability of the spin helical structure, the growth of the spin domains, and so on.  
Therefore, the hydrodynamics in multicomponent BECs is an active area of study.

In multicomponent BECs, novel turbulence not seen in scalar BECs can be realized; this turbulence is expected to exhibit  two types of statistical laws:
a statistical law characteristic of the system and  universal ones independent of the details of the system. 
The former law can  give  a new point of view to  turbulence studies, and the latter one enables us to study universal laws of turbulence. 
Therefore, turbulence in multicomponent BECs can offer opportunities to obtain novel viewpoints and study trends in turbulence. 
This is our motivation for the study of turbulence in multicomponent BECs. 

We have previously performed  theoretical and numerical studies of spin turbulence (ST) in spin-1 spinor BECs with a ferromagnetic (FM) interaction, 
where the spin density vector spatially points in various directions \cite{FT12a,FT12b, AT13}. 
In  ST, we focused on the spectrum of the spin-dependent interaction energy, finding theoretically and numerically the $-7/3$ power law.   
In CT and QT, the Kolmogorov $-5/3$ power law is known to appear in the kinetic energy, but we found a novel $-7/3$ power law in the spectrum of 
the spin-dependent interaction energy in our previous work \cite{FT12a}. These studies of ST were performed in the spin-1 spinor BEC with a FM interaction, but we have not investigated  ST with an antiferromagnetic (AFM) interaction in detail; in such a case the spin magnitude is small. 
This ST is expected to exhibit  behavior much different from  ST with a FM interaction.

In this paper, we report the  characteristic properties of  ST with small spin magnitude in spin-1 spinor BECs that arises because of the counterflow instability \cite{FT12a}. 
This kind of ST is obtained when the spin-dependent interaction is AFM or a static magnetic field is applied to the system with a FM interaction.
In the former case, the spin magnitude obviously becomes small because of the AFM interaction.
In the latter case, the quadratic Zeeman effect reduces the magnitude of the spin density vector. 
In such ST, we find that the $-1$ and $-7/3$ power laws can appear in the low- and high-wave-number regions by using a Kolmogorov-type dimensional scaling analysis for 
the equations of motion of the spin vector and the nematic tensor.
These power laws are investigated by  numerically calculating the spin-1 spinor GP equations.  

\section{Formulation} 
\subsection{Spinor Gross-Pitaevskii equations}  
We consider a spin-1 spinor BEC at zero temperature under a magnetic field in the $z$ direction, 
which is well described by the macroscopic wave functions $\psi _m$ ($m = 1,0,-1$) with the magnetic quantum number $m$. 
 The wave functions $\psi _m$ obey the spinor GP equations \cite{Ohmi98, Ho98}
 \begin{eqnarray}
 i\hbar \frac{\partial}{\partial t} \psi _{m}  &=&  \left(-\frac{\hbar ^2 }{2M} \nabla ^2 + V - pm + qm^{2}\right) \psi _{m} \nonumber \\
 &+& c_{0} \rho \psi _{m} + c_{1}  \bm{F} \cdot {\hat{\bm{F}} } _{mn} \psi _{n}.  \label{GP}
 \end{eqnarray}
In this paper, Greek indices that appear twice are to be summed over $x$, $y$, $z$, and Roman indices are to be summed over $-1, 0, 1$.

The first four terms on the right-hand side of Eq. (\ref{GP}) comprise the single-particle part, 
which contains the kinetic, potential, linear, and quadratic Zeeman terms. 
The kinetic and potential terms, where $M$ and $V$ are the mass of a particle and the trapping potential, respectively, are the same as the scalar GP equation. 
The remaining two terms with  coefficients $p$ and $q$ are the linear and quadratic Zeeman terms. 
The former term leads to  Larmor spin precession, whereas the latter one decreases (increases) the magnitude of the $z$ component of 
the spin density vector  for positive (negative) $q$.    

The interaction part is composed of  spin-independent and spin-dependent interactions with coefficients $c_{0}$ and $c_{1}$, which are 
expressed by $4 \pi \hbar ^{2} (a_{0} + 2a_{2})/3M$ and $4 \pi \hbar ^{2} (a_{2}  - a_{0})/3M$. Here, $a_{0}$ and $a_{2}$ are the $s$-wave scattering lengths corresponding to the total spin-$0$ and spin-2 channels. 
The total density $\rho$ and the spin density vector $F_{\mu}$ ($\mu = x, y, z$ ) are given by $\rho =  |\psi _m|^2$ and  $F_{\mu} =  \psi _{m}^{*} ({\hat F}_{\mu})_{mn} \psi _{n}$, where $({\hat F}_{\mu})_{mn}$ are the spin-1 matrices.
The spin-independent interaction is similar to that in the binary GP equations, which conserves the particle number of each component. 
In contrast, the spin-dependent interaction is characteristic of spinor BECs, exchanging  particles among the different components. 
The spin-dependent interaction energy $E_{s}$ is given by 
 \begin{eqnarray}
E_{s} = \frac{c_{1}}{2} \int \bm{F} ^{2} d \bm{r}. \label{spin_energy}
 \end{eqnarray}
This expression shows that, because $c_{1}$ is positive (negative), the spin-dependent interaction is AFM (FM), which is very important for the dynamics of the spin density vector in spinor BECs. 
 
\subsection{Continuity equations for spin vector and nematic tensor} 
In this section, we describe the continuity equations for the spin vector and the nematic tensor, which are used to derive 
the $-1$ and $-7/3$ power laws in the spectrum of the spin-dependent interaction energy in Sec. III. 
These equations are discussed by Yukawa and Ueda \cite{YU12}. 

We introduce the spin vector and the nematic tensor defined by 
\begin{eqnarray}
f_{\mu} = \frac{1}{\rho}  \psi _{m}^{*} (\hat{F}_{\mu})_{mn} \psi _{n} , \label{spin}
\end{eqnarray}
\begin{eqnarray}
n_{\mu \nu} = \frac{1}{\rho} \psi _{m}^{*} (\hat{N}_{\mu \nu})_{mn} \psi _{n} \label{nematic}
,\end{eqnarray}
respectively, with 
\begin{eqnarray}
(\hat{N}_{\mu \nu})_{mn}  = \frac{1}{2} [ (\hat{F}_{\mu})_{ml} (\hat{F}_{\nu})_{ln} + (\hat{F}_{\nu})_{ml} (\hat{F}_{\mu})_{ln} ]. 
\end{eqnarray}

The continuity equations for the spin vector and the nematic tensor are derived from the spin-1 spinor GP equations (\ref{GP}). 
The continuity equation for the spin vector is 
\begin{eqnarray}
\frac{\partial}{\partial t} \rho f_{\mu} +  \nabla \cdot \rho \bm{v}_{\mu} = \frac{1}{\hbar} \epsilon _{z \mu \nu}  \rho (p  f_{\nu} - 2q  n_{z \nu}),  \label{spin continuity}
\end{eqnarray}
where the spin current is defined by 
\begin{eqnarray}
\bm{v}_{\mu} = f_{\mu} \bm{v} - \frac{\hbar}{M} \epsilon _{\mu \nu \lambda} \left[ \frac{1}{4} f_{\nu} (\nabla f_{\lambda}) + n_{\nu \eta} (\nabla n _{\lambda \eta})\right].\label{spin_current}
\end{eqnarray}
Here
\begin{eqnarray}
 \bm{v} = \frac{\hbar}{2M\rho i}  [ \psi _{m} ^{*} (\nabla \psi _{m}) - (\nabla \psi _{m} ^{*}) \psi _{m}] 
\end{eqnarray}
is the superfluid velocity.
Similarly, the continuity equation for the nematic tensor is obtained from   
\begin{eqnarray}
\frac{\partial}{\partial t} \rho n_{\mu \nu} + \nabla \cdot \rho \bm{v}_{\mu \nu} &=& \frac{\rho}{\hbar}\Big[  \epsilon _{z \mu \lambda}\left( p n_{\nu \lambda} - \frac{q}{2} \delta _{z \nu} f_{\lambda}\right)  \nonumber \\
&+&  \epsilon _{z \nu \lambda}\left( p n_{\mu \lambda} - \frac{q}{2} \delta _{z \mu} f_{\lambda}\right)  \Big] \nonumber  \\
&+& \frac{c_{1} \rho^{2}}{\hbar} ( \epsilon _{\mu \lambda \eta} f_{\lambda} n _{\nu \eta} + \epsilon _{\nu \lambda \eta} f_{\lambda} n _{\mu \eta}  ), \label{nematic continuity}
\end{eqnarray}
where the nematic tensor current is defined by
\begin{eqnarray}
\bm{v}_{\mu \nu} = n_{\mu \nu} \bm{v} - \frac{\hbar}{4M}  \{ \epsilon _{\mu \lambda \eta}[ f_{\lambda} (\nabla n_{\nu \eta}) - (\nabla f_{\lambda}) n_{\nu \eta}]  \nonumber \\
+ \epsilon _{\nu \lambda \eta}[ f_{\lambda} (\nabla n_{\mu \eta}) - (\nabla f_{\lambda}) n_{\mu \eta}] \}. \label{nematic current}
\end{eqnarray}
In \cite{YU12}, the equation of motion for the superfluid velocity is also derived, but we do not show the expression because this equation is not 
relevant to the derivation of the $-1$ and $-7/3$ power laws. 

Finally, we note that there are some constraint conditions for the equivalence between the spin-1 spinor GP equations 
and the hydrodynamic equations:  
 \begin{eqnarray}
n _{\mu \mu} = 2, \label{constraint1}
\end{eqnarray}
 \begin{eqnarray}
n_{\mu \nu} f_{\nu}  = f_{\mu}, \label{constraint2}
\end{eqnarray}
 \begin{eqnarray}
{\rm det} \hspace{1mm} n _{\mu \nu} =  \frac{1}{4} f_{\mu}^{2}. \label{constraint3}
\end{eqnarray}
These constraint conditions are discussed in \cite{YU12}. 

\section{Derivation of the $-1$ and $-7/3$ power laws}
We find that the $-1$ and $-7/3$ power laws in the spectrum of spin-dependent interaction energy can appear in  ST 
when the magnitude of the spin vector is small. 
In this section, we treat a uniform system without a magnetic field and describe the derivation of two power laws using almost the same as 
method used in previous studies \cite{scaling1,scaling2,FT12a}. 

\subsection{Spectrum of spin-dependent interaction energy} 
We show an expression for the spectrum of the spin-dependent interaction energy.
The spin-dependent interaction energy $\mathcal{E} _{s}$ per unit volume is given by 
\begin{equation}
\mathcal{E} _{s} = \frac{c_{1}}{2L^{n_{d}}} \int \bm{F}(\bm{r})^{2} d\bm{r},
\end{equation}
where $L$ and $n_{d}$ are the system size and the space dimension, respectively. 
We expand the spin density vector $\bm{F}(\bm{r})$ with  plane waves as
$\bm{F}(\bm{r}) = \sum _{\bm{k}} \tilde{\bm{F}}(\bm{k}) e^{i\bm{k}\cdot\bm{r}}$.
Then the spin-dependent interaction energy $E_{s}$ is represented by $\tilde{\bm{F}}(\bm{k})$ as
$\mathcal{E} _{s} = \frac{c_{1}}{2} \sum _{\bm{k}} |\tilde{\bm{F}}(\bm{k})|^{2}$. 
Therefore, the energy spectrum of the spin-dependent interaction energy is given by
\begin{eqnarray}
\mathcal{E} _{s} (k) &=& \frac{c_{1}}{2 \Delta k} \sum _{k<|\bm{k}_{1}|<k+\Delta k} |\tilde{\bm{F}}(\bm{k}_{1})|^{2} \nonumber \\  
&=& \frac{\pm 1 }{2 \Delta k} \sum _{k<|\bm{k}_{1}|<k+\Delta k} |\tilde{\bm{A}}(\bm{k}_{1})|^{2}, \label{spectrum} 
\end{eqnarray}
where $\Delta k$ and $A_{\mu}$ are $2\pi /L$ and $\sqrt{|c_{1}|} F_{\mu}$, respectively.
The  $+$ and $-$ signs denote whether the spin-dependent interaction is AFM or FM, respectively.

\subsection{Kolmogorov-type dimensional scaling analysis}
As  preparation for the derivation of the $-1$ and $-7/3$ power laws, we briefly review CT in three-dimensional systems,  in which vortices are considered to be important for  understanding  the kinetic energy spectrum. 
In this turbulence, external forces generate  large vortices, which reconnect with each other, and smaller vortices are nucleated. 
Furthermore, these small vortices also reconnect with each other, thus splitting up into even smaller vortices. 
This reconnection of vortices can occur until the size of the vortex is comparable to 
the Kolmogorov scale, below which the viscosity is dominant and the kinetic energy dissipates. As a result, the vortices disappear on this scale.
The wave number region where the reconnections of vortices make smaller ones without dissipation is called the inertial range.  
In this region, the kinetic energy seems to be constantly transferred from the low to  high wave numbers, which means 
the existence of a constant kinetic energy flux independent of the wave number. 
This constant energy flux leads to the Kolmogorov $-5/3$ power law in the kinetic energy spectrum, which is 
confirmed by many numerical and experimental studies \cite{Davidson,Frisch}. 
In the current derivation of the $-1$ and $-7/3$ power laws, we apply this assumption for the constant energy flux to ST. 

We apply three approximations to Eqs. (\ref{spin continuity})--(\ref{nematic current}) to obtain the equations for the derivation of $-1$ and $-7/3$ power laws. 

The first approximation is that the total density $\rho$ is uniform: $\rho (\bm{r}) \sim \rho _{0} = N/L^{2}$ with  total particle number $N$ and
system size $L$. This is valid for $|c_{0}/c_{1}| \gg 1$, which is satisfied in the usual experiments. Thus, in Eqs. (\ref{spin continuity})--(\ref{nematic current}), we can neglect the spatial derivative of total density \cite{density}. 

The second approximation is that the superfluid velocity is much smaller than the sound velocity. 
In our numerical calculations described in Sec. IV, immediately after the instability occurs, many vortices can be nucleated. 
They can then induce a superfluid velocity comparable to the sound velocity $C_{s} = \sqrt{c_{0} \rho _{0} / 2M  }$ near the vortex core. 
However, as  ST is formed, the vortices can disappear via pair annihilation. 
Therefore, we consider that the superfluid velocity is much smaller almost everywhere than the sound velocity in ST and can neglect the terms with  superfluid velocity 
in Eqs. (\ref{spin continuity})--(\ref{nematic current}). 
Further, even if there are vortices, the velocity can be comparable to the sound velocity only near the vortex core, 
which means that the vortices do not affect 
the spectrum in the wave number region lower than $k_{s}$ corresponding to the spin coherence length 
$\xi _{s} = \hbar / \sqrt{2M |c_{1}| \rho_{0}}$. In our previous study \cite{FT12a}, we used the same approximations. 

The third approximation is that the magnitude of the spin vector is smaller than unity. 
This is valid in a system with  an AFM interaction or a FM interaction under a static magnetic field because 
the AFM interaction or the quadratic Zeeman effect reduces the magnitude of the spin vector. 
Thus, by using this approximation,  we can neglect the term with the spin vector in Eq. (\ref{spin_current}) because the nematic tensor and the spin vector are related by the relation $n_{\mu \nu}^{2} = -\frac{1}{2}f_{\mu}^{2} + 2$ \cite{Large_spin}. 

Applying these three approximations to Eqs. (\ref{spin continuity})--(\ref{nematic current}), 
we obtain the following equations:
\begin{eqnarray}
\frac{\partial}{\partial t} f_{\mu} +  \nabla \cdot \bm{v}_{\mu} = 0,  \label{spin continuity_s}
\end{eqnarray}
\begin{eqnarray}
\bm{v}_{\mu} = - \frac{\hbar}{M} \epsilon _{\mu \nu \lambda} n_{\nu \eta} (\nabla n _{\lambda \eta}),\label{spin_current_s}
\end{eqnarray}
\begin{eqnarray}
\frac{\partial}{\partial t} n_{\mu \nu} + \nabla \cdot \rho \bm{v}_{\mu \nu} = \frac{c_{1} \rho}{\hbar} ( \epsilon _{\mu \lambda \eta} f_{\lambda} n _{\nu \eta} + \epsilon _{\nu \lambda \eta} f_{\lambda} n _{\mu \eta}  ), \label{nematic continuity_s}
\end{eqnarray}
\begin{eqnarray}
\bm{v}_{\mu \nu} = - \frac{\hbar}{4M}  \{ \epsilon _{\mu \lambda \eta}[ f_{\lambda} (\nabla n_{\nu \eta}) - (\nabla f_{\lambda}) n_{\nu \eta}]  \nonumber \\
+ \epsilon _{\nu \lambda \eta}[ f_{\lambda} (\nabla n_{\mu \eta}) - (\nabla f_{\lambda}) n_{\mu \eta}] \} \label{nematic current_s}
.\end{eqnarray}

In the following, we apply a Kolmogorov-type dimensional scaling analysis to Eqs. (\ref{spin continuity_s})--(\ref{nematic current_s}), where 
the scale transformation is separately performed in the low- ($k < k_{b}$) and high- ($ k_{b} < k $) wave-number regions. 
Here the boundary wave number $k_{b} = 2 \sqrt{|c_{1}| M \rho _{0}}/\hbar$ is obtained by the condition in which the second term on the left-hand side of 
Eq. (\ref{nematic continuity_s}) and the term on the right-hand side are comparable.
It is not obvious whether the scale transformation can be separately performed, which is explained in Sec. V. A. 

We comment on the physical meaning of the boundary wave number $k_{b}$, which decides whether the dispersion relation of the spin wave in the polar phase becomes phonon-like or free-particle-like. 
In the system with an AFM interaction, the ground state is polar phase. By solving the Bogoliubov-de Gennes equation with this phase, the dispersion relation of the spin wave 
is given by $\hbar \omega = \sqrt{\hbar \omega _{0}(\hbar \omega _{0} + 2c_{1}\rho _{0})} $ with $\hbar \omega  _{0} = \hbar^{2}k^{2}/2M$, which shows that 
the dispersion relation is proportional to $k$ and $k^{2}$ in the low- ($k < k_{b}$) and high- ($ k_{b} < k $) wave-number regions, respectively.

First, we consider the low-wave-number region $ k < k_{b}$.   
In this region, the second term on the left-hand side of Eq. (\ref{nematic continuity_s}) can be neglected, so that Eq. (\ref{nematic current_s}) 
is not necessary in the following.
Then, the scaling analysis is applicable to the remaining terms in Eqs. (\ref{spin continuity_s})--(\ref{nematic continuity_s}).  
We perform the scale transformation $\bm{r}\rightarrow \alpha \bm{r}$ and $ t \rightarrow \beta t$ in Eqs. (\ref{spin continuity_s})--(\ref{nematic continuity_s}). 
Then, if $f_{\mu}$ and $n _{\mu \nu}$ are transformed to 
$f_{\mu} \rightarrow \beta^{-1} f_{\mu}$ and $n_{\mu \nu} \rightarrow \alpha \beta^{-1} n_{\mu \nu}$, 
Eqs. (\ref{spin continuity_s})--(\ref{nematic continuity_s}) are invariant. Thus, in the low-wave-number region $k < k_{b}$, we obtain
\begin{eqnarray}
f_{\mu} \sim C_{L} t^{-1} , \label{spin_transformation1}
\end{eqnarray}
where $C_{L}$ is a dimensional constant. Then, $A_{\mu}$ is expressed by 
\begin{eqnarray}
A_{\mu} \sim \Lambda_{L} t^{-1}  \label{spin_transformation12}
\end{eqnarray}
with a dimensional constant $\Lambda _{L} = \sqrt{|c_{1}|}C_{L}$.
Also, in ST, we suppose that the energy flux $\epsilon_{L}$ of the spin-dependent interaction energy in the wave number space is independent of the wave number. This assumption is equivalent to the existence of a wave number region in which the energy is constantly transferred. 
Therefore, in ST with small spin magnitude, the spectrum of the spin-dependent interaction energy in the low-wave-number region should be dominated 
by $\epsilon _{L}$ and $\Lambda_{L}$.
Then, by  dimensional analysis, the relation between the characteristic time $t_{s}$ and the energy flux $\epsilon_{L}$ is given by
 \begin{eqnarray}
\epsilon _{L} \sim \frac{A_{\mu}^{2}}{t_{s}} \sim \Lambda_{L}^{2} t_{s}^{-3}. \label{energy_transport1}
\end{eqnarray}
Using Eqs. (\ref{spectrum}), (\ref{spin_transformation12}), and (\ref{energy_transport1}), we obtain the $-1$ power law in the low-wave-number region $k < k_{b}$ 
by  dimensional analysis:  
 \begin{eqnarray}
|\mathcal{E} _{s} (k)| &\sim& \frac{A_{\mu}^{2}}{k}  \\ \nonumber
               &\sim&  \Lambda_{L}^{2} k^{-1} t_{s}^{-2} \sim \epsilon _{L} ^{2/3} \Lambda_{L}^{2/3} k^{-1}. \label{-1_power_law}
\end{eqnarray}

In contrast, in the high-wave-number region $ k_{b} < k$, the term on the right-hand side of Eq. (\ref{nematic continuity_s}) can be neglected.
Then, if the spin vector $f_{\mu}$ and $n _{\mu \nu}$ are transformed to $f_{\mu} \rightarrow \alpha ^{2} \beta^{-1} f_{\mu}$ and $n_{\mu \nu} \rightarrow \alpha^{2} \beta^{-1} n_{\mu \nu}$, Eqs. (\ref{spin continuity_s})--(\ref{nematic current_s}) are invariant 
under the scale transformation. Thus, we obtain
\begin{eqnarray}
A_{\mu} \sim \Lambda_{H} r^{2}t^{-1}  \label{spin_transformation2}
\end{eqnarray}
with a dimensional constant $\Lambda _{H}$.
In the same way as in the above argument, the spectrum in the high-wave-number region should be dominated by a constant energy flux $\epsilon _{H}$ and 
a dimensional constant $\Lambda_{H}$, 
which leads to the $-7/3$ power law in the high-wave-number region $k_{b} < k$:
 \begin{eqnarray}
|\mathcal{E} _{s} (k)|&\sim& \frac{A_{\mu}^{2}}{k}  \\ \nonumber
               &\sim& \Lambda_{H}^{2} k^{-5}  t_{s}^{-2} \sim \epsilon _{H} ^{2/3} \Lambda_{H}^{2/3} k^{-7/3}, \label{-7/3_power_law}
\end{eqnarray}
where $\epsilon _{H}$ is given by
 \begin{eqnarray}
\epsilon _{H} \sim \frac{A_{\mu}^{2}}{t_{s}} \sim  \Lambda_{H}^{2} k^{-4} t_{s}^{-3}. \label{energy_transports}
\end{eqnarray}

We note the scaling regions with the $-1$ and $-7/3$ power laws. The above derivation of two power laws shows that these laws can appear 
in the low- ($k < k_{b}$) and high- ($ k_{b} < k$) wave-number regions, respectively. However, the spectrum must be affected by the structure of the spin vortices or spin 
domain walls in the region $ k_{s} = 2 \pi /\xi _{s} < k$ \cite{defect}. Thus, the spectrum should exhibit a $-7/3$ power law in the region $k_{b} < k  < k_{s}$. 
In contrast, the $-1$ power law should appear in the low-wave-number region $k_{L} < k < k_{b}$, 
where $k_{L} = 2 \pi /L$ is the wave number corresponding to the system size $L$. However, the boundary condition may affect the spectrum near $k_{L}$.

Finally, we comment on the spectrum at the boundary wave number $k_{b}$. 
In the vicinity of $k_{b}$, the above approximations for Eq. (\ref{nematic continuity_s}) are invalid, 
so that we cannot estimate the scale transformations of the spin vector and the nematic tensor. At present, the spectrum near $k_{b}$ cannot be found 
from the scaling analysis; this   will a subject of future  study. 

Summarizing our results, we find that, in  ST with  small spin magnitude, the spectrum the of spin-dependent interaction energy can exhibit $-1$ and $-7/3$ power laws in the low- ($k_{L} < k < k_{b}$) and high- ($ k_{b} < k  < k_{s}$) wave-number regions, respectively.

\section{Numerical Results}  
We show  numerical results for  ST with small spin magnitude in a two-dimensional uniform system, which is realized in  two cases: (i) with an AFM interaction without a magnetic field 
and (ii)  with a FM interaction under a static magnetic field. In case (i), the AFM interaction reduces the magnitude of the spin vector, whereas, in  case (ii), the 
quadratic Zeeman effect accomplishes this reduction. 
Therefore, the $-1$ and $-7/3$ power laws are expected to appear in both cases. 

\begin{figure} [t]
\begin{center}
\includegraphics[width=80mm]{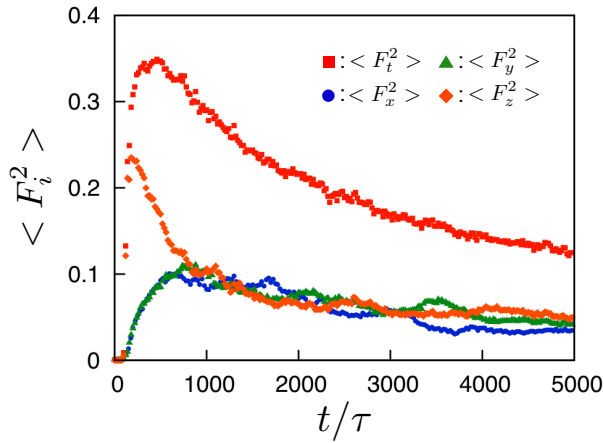}
\caption{(Color online) Time dependence of the spatial average for the squared magnitude of the normalized spin density vector in a uniform system with an AFM interaction. 
The quantities $\left\langle F_{i}^{2}\right\rangle$ ($i = x, y, z, t$) are defined by Eqs. (\ref{average_squared_magnitude1}) and (\ref{average_squared_magnitude2}). 
As the counterflow instability occurs, the magnitude of the spin density vector begins to rapidly grow at $t/\tau \sim 90$. After the instability, the magnitude monotonically decreases because of the AFM interaction. }
\label{fig1}
\end{center}
\end{figure}

\begin{figure} [!b]
\begin{center}
\includegraphics[width=80mm]{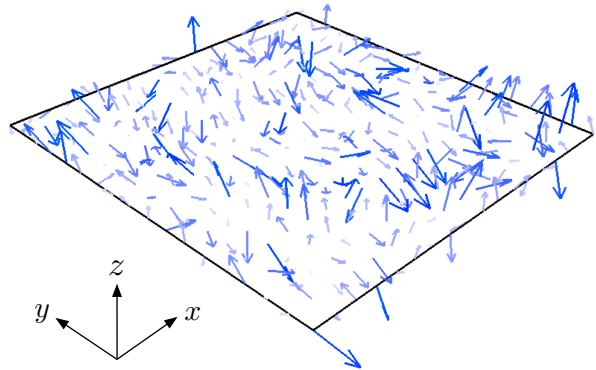}
\caption{(Color online)   Distribution of spin density vector $\bm{F}$ at $t/\tau = 3500$ in Fig. 1. The system size $L \times L$ is $256 \xi \times 256 \xi$.
The spin density vector points in  various directions, and  ST is realized. }
\label{fig2}
\end{center}
\end{figure}

\begin{figure*} [!t]
\begin{center}
\includegraphics[width=180mm]{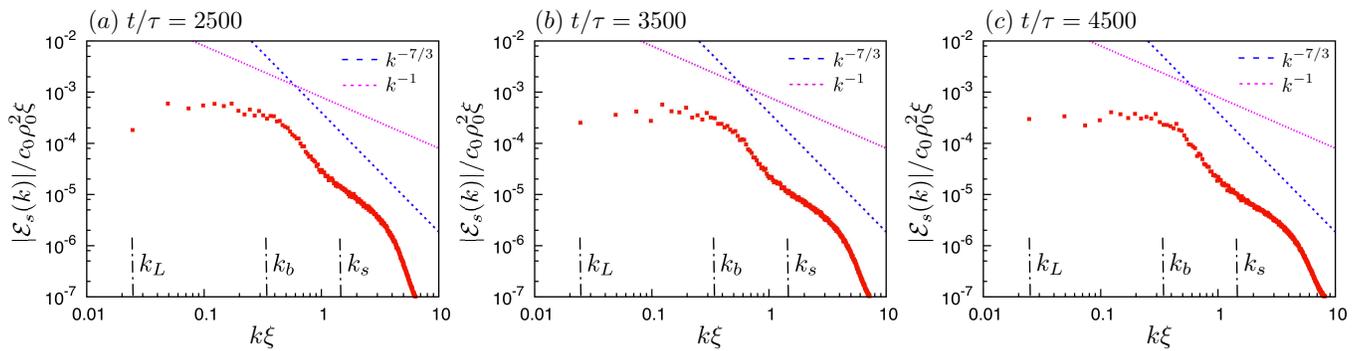}
\caption{(Color online)  Time development of the spectrum of the spin-dependent interaction energy in a uniform system with an AFM interaction. The spectra at (a) $t/\tau = 2500$, (b) $t/\tau = 3500,$ and  (c) $t/\tau = 4500$ are shown in log-log coordinates. The dotted and fine dotted lines are proportional to $k^{-7/3}$ and $k^{-1}$, respectively. The expressions for $k_{L}$, $k_{b}$, and $k_{s}$ are given in Sec. III.
The $-7/3$ power law appears in the high-wave-number region $k_{b}<k<k_{s}$, whereas the spectrum near $k_{L}$ largely deviates from the $-1$ power law. }
\label{fig3}
\end{center}
\end{figure*}

ST in a uniform system can be realized by the counterflow instability, in which a spatial density modulation with a stripe structure 
is induced and the collapse of the structure leads to the ST \cite{FT12a}.  All ST cases in this paper are obtained by the counterflow instability.

\begin{figure} [!b]
\begin{center}
\includegraphics[width=80mm]{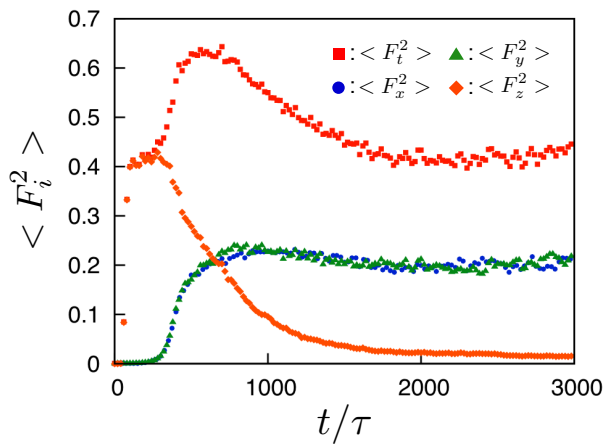}
\caption{(Color online) Time dependence of the spatial average for the squared magnitude of the normalized spin density vector in a uniform system 
with a FM interaction under a static magnetic field. 
The quantities $\langle F_{i}^{2}\rangle$ ($i = x, y, z, t$) are defined by Eqs. (\ref{average_squared_magnitude1}) and (\ref{average_squared_magnitude2}). 
The counterflow instability occurs, leading to  rapid growth of the magnitude of the $z$ component at $t / \tau \sim 50$. After a while, 
 the $x$ and $y$ components grow at $t / \tau \sim 300$. Then, the magnitude of the $z$ component decreases because of the quadratic Zeeman effect.  
 As a result, $\langle F_{t}^{2}\rangle$ becomes small in spite of the FM interaction.}
\label{fig4}
\end{center}
\end{figure}

\begin{figure} [!b]
\begin{center}
\includegraphics[width=80mm]{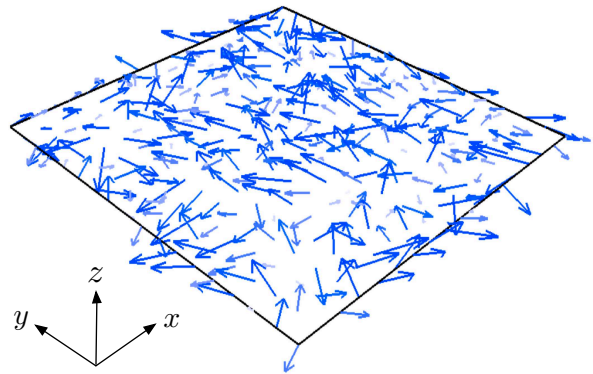}
\caption{(Color online)   Distribution of spin density vector $\bm{F}$ at $t/\tau = 2000$ in Fig. 4. The system size $L \times L$ is $256 \xi \times 256 \xi$. 
The spin density vector lies on the $x$-$y$ plane because the 
quadratic Zeeman effect reduces the $z$ component of the spin density vector. }
\label{fig5}
\end{center}
\end{figure} 

\subsection{ST with the AFM interaction}
We briefly present the parameters and the initial state for the numerical calculation. 
Our system is assumed to be uniform, so that the potential $V$ is zero everywhere.
The system size $L \times L$ is $256 \xi \times 256 \xi$ with a coherence length $\xi = \hbar \sqrt{2Mc_{0} \rho_{0}}$.
Here, $\rho_{0}$ is the initial total density, which is given by $N/L^{2}$ with  total particle number $N$. 
The ratio of the interaction parameters, $|c_{0}/c_{1}|$, is $20$, 
where $c_{0}$ and $c_{1}$ are positive. The relative velocity $V_{R}$ between the $m = \pm 1$ components is $1.178 C_{s}$. 
The initial state $\psi _{m} $ ($m = 1, 0,-1$) for the counterflow between the $m=\pm 1$ components is expressed by 
\begin{equation}
\begin{pmatrix} 
\psi _{1} \\
\psi _{0} \\
\psi _{-1}
\end{pmatrix}
= \sqrt{\frac{\rho _{0}}{2}}
\begin{pmatrix} 
{\rm{exp}}(i\frac{MV_{R}}{2 \hbar } x ) \\
0 \\
{\rm{exp}}(-i\frac{MV_{R}}{2 \hbar} x)
\end{pmatrix}
. 
\end{equation}
In this state, the $m=1$ component moves in the $x$ direction, whereas the $m=-1$ component moves in the opposite direction; this induces the counterflow instability and leads to ST. We add some small white noise to the initial state of Eq. (27). 
The details of the counterflow instability are described in \cite{FT12a}.

\begin{figure*} [!t]
\begin{center}
\includegraphics[width=180mm]{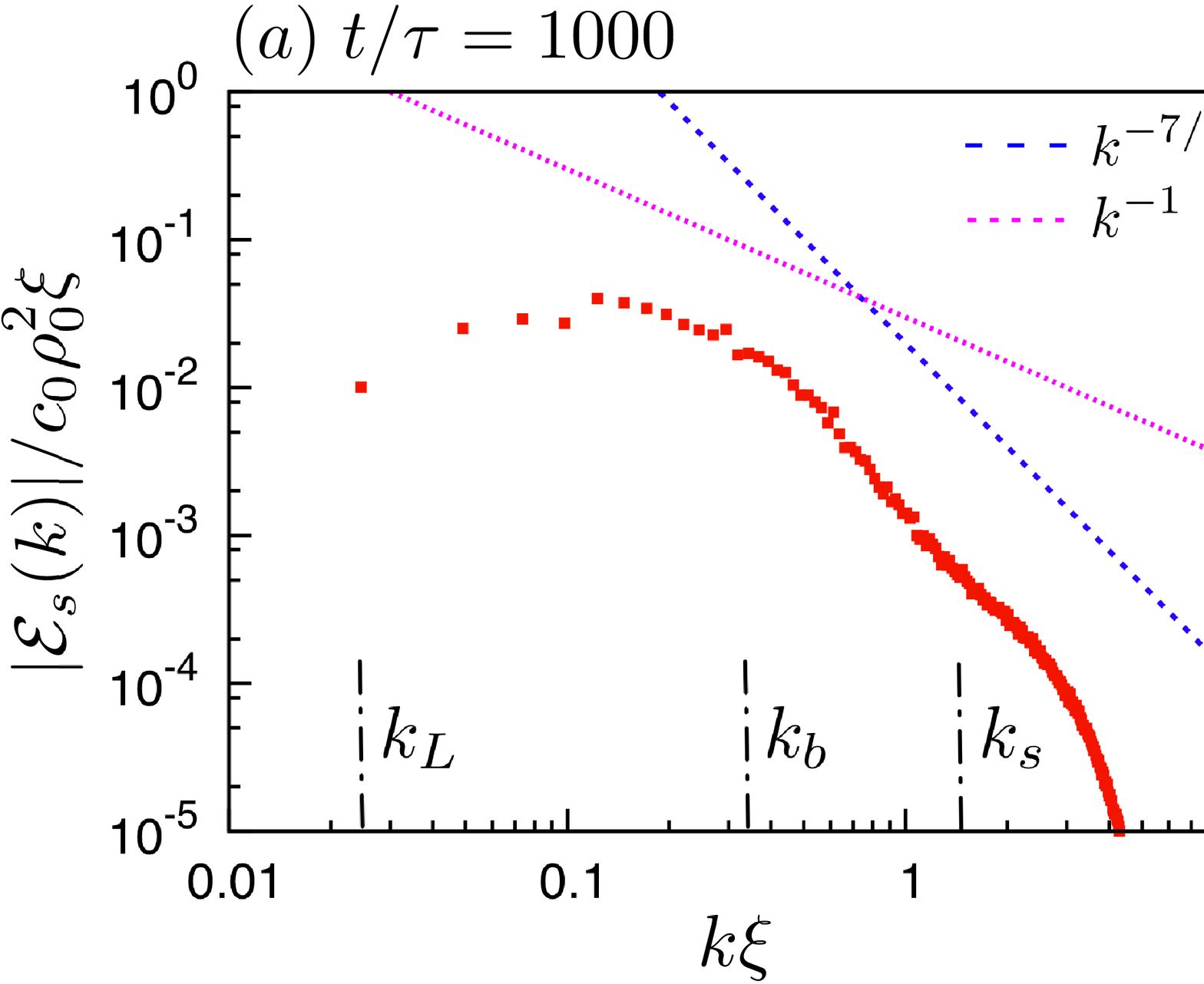}
\caption{(Color online)  Time development of the spectrum of the spin-dependent interaction energy in a uniform system with a FM interaction under a static magnetic field. The spectra at (a) $t/\tau = 1000$, (b) $t/\tau = 2000,$ and (c) $t/\tau = 3000$ are shown in log-log coordinates. The dotted and fine dotted lines are proportional to $k^{-7/3}$ and $k^{-1}$, respectively. The expressions for $k_{L}$, $k_{b}$, and $k_{s}$ are given in Sec. III. The $-1$ and $-7/3$ power laws appear clearly in (c). }
\label{fig6}
\end{center}
\end{figure*}

Figure 1 shows the time dependence of the spatial average for the squared magnitude of the normalized spin density vector, which is defined by 
 \begin{eqnarray}
\left\langle F_{i}^{2}\right\rangle = \frac{1}{\rho _{0}^{2}L^{2}} \int F_{i}^{2}( \bm{r}) d\bm{r} \hspace{5mm} (i = x, y, z)  \label{average_squared_magnitude1}
\end{eqnarray}
and
 \begin{eqnarray}
\left\langle F_{t}\right\rangle^{2} = \sum _{i = x,y,z} \left\langle F_{i}^{2}\right\rangle . \label{average_squared_magnitude2}
\end{eqnarray}
In the initial state, the magnitude of the spin density vector is almost zero because the  $m = \pm 1$ components are spatially miscible. 
However, as the counterflow instability rapidly grows at $t/\tau \sim 90$, the magnitude increases, as shown in Fig. \ref{fig1}. 
After the onset of the counterflow instability,  the spin density vector spatially points in various directions and  ST is realized, as shown in Fig. \ref{fig2}. 
Then, the magnitude $\left\langle F_{i}^{2}\right\rangle$ monotonically decreases because of the AFM interaction, which is qualitatively different from  ST with a FM interaction. 

We numerically calculate the spectrum of the spin-dependent interaction energy, as shown in Fig. \ref{fig3}. The spectrum exhibits the expected $-7/3$ power law in the high-wave-number region $k_{b}<k<k_{s}$, but it deviates from the $-1$ power in the low-wave-number region $k_{L}<k<k_{b}$. 
After the onset of the counterflow instability, in the spectrum of Fig.  \ref{fig3}(a), the $-7/3$ power law appears in the high-wave-number region and, in the low-wave-number region, the spectrum exhibits the sign of the 
$-1$ power law. As  time progresses, the $-1$ power law appears 
near the boundary wave number $k_{b}$ in Fig.  \ref{fig3}(b). 
However, at  sufficiently longer times, the $-1$ power just declines, as shown in Fig. \ref{fig3}(c), because the AFM interaction 
reduces the spin-dependent interaction energy. 
Thus, the spectrum in the low-wave-number region is considered to have difficulty in growing, which may disturb the $-1$ power law. 

\subsection{ST with the FM interaction under a static magnetic field}

We show the numerical results for  ST with a FM interaction under a static magnetic field. In this system, the quadratic Zeeman effect reduces the magnitude of the spin density vector, which is expected to lead to the $-1$ and $-7/3$ power laws in the spectrum of the spin-dependent interaction energy. 

This numerical calculation is almost the same as that for the case of Sec. IV A, but there are two differences. 
The first difference is the sign of the spin-dependent interaction. In the case for the FM interaction, $c_{1}$ is negative, so that we use the parameter $c_{1}/c_{0} = -20$ with positive $c_{0}$. The second difference is the application of a static magnetic filed, whose effect is included in Eq. (\ref{GP}) as the linear and quadratic Zeeman terms. 
In our numerical calculation, we omit the linear Zeeman effect because this effect only induces  Larmor spin precession, which does not affect the spectrum. Thus, we use the parameters $p = 0$ and $q = 1.2|c_{1}| \rho _{0}$. The strength of the magnetic field is discussed in Sec. V C.

Figure \ref{fig4} shows the time dependence of the spatial average for the squared magnitude of the normalized spin density vector. The counterflow instability occurs at $t/\tau = 50$, which 
causes the $z$ component of the spin density vector to grow rapidly. 
Slightly after the onset of the instability, the $x$ and $y$ components grow belatedly. As  time progresses, the $z$ component decreases because of the quadratic Zeeman 
effect. The  spin density vector distribution at $t/\tau = 2000$ is shown in Fig. \ref{fig5}, where one can see that the spin density vector lies almost on the $x$-$y$ plane. 
At this time, the magnitude of the spin density vector becomes small. 

Figures \ref{fig1} and \ref{fig4} show the difference between  ST with an AFM interaction and that with a FM interaction under 
a static magnetic field; $\langle F_{t}^{2}\rangle$ in the former case is smaller than that in the latter one. 
In the latter case, the effect of the quadratic Zeeman term can compete with that of the FM interaction to cause this difference.

The spectrum of the spin-dependent interaction energy clearly exhibits both the $-1$ and $-7/3$ power laws, as shown in Fig. \ref{fig6}. 
The $-7/3$ power law appears in Figs. \ref{fig6}(a)--\ref{fig6}(c), which is the same as the case for  ST with an AFM interaction. 
For the $-1$ power law, its sign appears slightly below the boundary wave number $k_{b}$ in Fig. \ref{fig6}(a). As the time passes, 
the spectrum near $k_{L}$ gradually grows in Fig. \ref{fig6}(b). Finally, as shown in Fig. \ref{fig6}(c),  the spectrum exhibits the $-1$ power law in the low-wave-number region $k_{L} < k < k_{b}$. 

This result is different from that of  ST with an AFM interaction. This difference is considered to be caused by the spin-dependent interaction. In a system with 
an AFM interaction, it seems to be difficult for the spectrum near $k_{L}$ to grow because this interaction only reduces the 
magnitude of the spin density vector. In contrast, the FM spin-dependent interaction tends to
increase the magnitude in the ST, enabling  the spectrum near $k_{L}$ to grow. 

In summary, in  ST with an AFM interaction in a uniform system, we find that the $-7/3$ power law appears in the high-wave-number region, 
but the spectrum in the low-wave-number region, particularly near $k_{L}$, deviates from the $-1$ power law.  
However, in  ST with a FM interaction under a static magnetic field, we find that both $-1$ and $-7/3$ power laws obviously appear 
in the high- and low-wave-number regions.  

\section{Discussion}  
We discuss three topics for ST with small spin magnitude.
In Sec. \ref{secA}, we discuss the scaling analysis and constraint conditions. 
There are the three constraint conditions given by Eqs. (\ref{constraint1})--(\ref{constraint3}) for the equivalence between the spin-1 spinor GP equations and the spin hydrodynamic equations. We consider the question of whether the scaling analysis for the derivation of the $-1$ and $-7/3$ power laws is 
consistent with these constraint conditions.  
In Sec. \ref{secB}, the localness of the interaction in turbulence is addressed.
In the derivation of the $-1$ and $-7/3$ power laws, we assume the localness of the interaction, which has been previously 
studied in other systems \cite{local1, local2}. 
In Sec. \ref{secC}, the influence of the magnetic field on the $-1$ and $-7/3$ power laws is discussed.

\subsection{Scaling analysis and constraint conditions}\label{secA}
Whether the Kolmogorov-type dimensional scaling analysis in Sec. III B is consistent with the constraint conditions of Eqs. (\ref{constraint1})--(\ref{constraint3})  is not obvious because our scaling analysis consists of two transformations, which are 
$f_{\mu} \rightarrow \beta^{-1} f_{\mu}$ and $n_{\mu \nu} \rightarrow \alpha \beta^{-1} n_{\mu \nu}$ in the low-wave-number region and 
$f_{\mu} \rightarrow \alpha^{2} \beta^{-1} f_{\mu}$ and $n_{\mu \nu} \rightarrow \alpha^{2} \beta^{-1} n_{\mu \nu}$ in the high-wave-number region. 
The constraint conditions must be satisfied at arbitrary time and position. At present, we do not completely understand whether the constraint conditions simultaneously satisfy 
the two transformations in the high- and low-wave-number regions.   

\subsection{Localness of interaction for energy flux in ST}\label{secB}
The localness of the interaction for the energy flux is very important for the Kolmogorov-type dimensional scaling analysis.
In two-dimensional CT,  modification of the spectrum by the nonlocalness of the interaction has been discussed \cite{local1}. 
In this system,  direct enstrophy and inverse energy cascades occur, where the $-3$ and $-5/3$ power laws are expected in the kinetic energy spectra \cite{inverse}.
However, this interaction is nonlocal in the inertial range, which leads to a logarithmic correction to the $-3$ power law \cite{local1}.    
Recently, the localness in the Kelvin wave cascade has been discussed \cite{local2}. 
Thus, the configuration of the spectrum is affected by the localness of the interaction. 

In our ST, we must investigate whether the interaction is local, but we do not understand the localness in ST because the hydrodynamic equations of the spin-1 spinor GP equation is complex. This problem remains the future work.

\subsection{Influence of magnetic field on the $-1$ and $-7/3$ power laws}\label{secC}
In this section, we discuss the influence of the magnetic field on the $-1$ and $-7/3$ power laws. In Sec. IV B, we applied a static magnetic field to a system with a FM interaction, where the spin magnitude becomes small and the $-1$ and $-7/3$ power laws appear. 
However, these power laws can be affected by the strength of the magnetic field.  
 
 In Sec. IV B, the terms generating the $-1$ and $-7/3$ power laws are a few times larger than the quadratic Zeeman term by an order estimation. Thus, although the quadratic Zeeman term is neglected in the scaling analysis, it is not  small, which may change these power laws. The quadratic Zeeman term has no spatial derivative, which means that the term can affect the spectrum in a wave number region smaller than $k_{Z} = \sqrt{qM/ \hbar ^{2} n}$, where $n$ is the order of the sum $\sqrt{n_{\mu \nu} ^{2}}$ of the nematic tensor. This is obtained by an order estimation between the kinetic and quadratic Zeeman terms in Eq. (\ref{spin continuity}) or (\ref{nematic continuity}). If the quadratic Zeeman effect is dominant, the effect must appear in the wave number region lower than $k_{Z} \sim 0.13$. However, the spectrum clearly exhibits the $-1$ power law there in Fig. \ref{fig6}(c). Therefore, the Zeeman effect in Sec. IV B seems too small to affect the 
 appearance of the $-1$ power law. 
 
Let us consider the magnetic field smaller than that in the case of Sec. IV B. Then, the magnitude of the spin vector is larger than that in Sec. IV B 
because of the small quadratic Zeeman effect. 
This leads to a deviation from the $-1$ power law, which is confirmed by our numerical calculation. 
Thus, we cannot obtain a clear $-1$ power law when the magnetic field is extremely small. 
The details of the influence of the magnetic field on ST will be studied in a future work.

\section{Conclusion}  
We have studied  ST with small spin magnitude in a spin-1 spinor BEC in a uniform system by using the spin-1 spinor GP equations. 
The $-1$ and $-7/3$ power laws in the low- ($k_{L}<k<k_{b}$) and high- ($k_{b}<k<k_{s}$) wave-number regions are derived  
by a Kolmogorov-type dimensional scaling analysis. 
We perform  numerical calculations for a two-dimensional uniform system that  
show that  ST with an AFM interaction exhibits the $-7/3$ power law, 
but the spectrum deviates from the $-1$ power law in the low-wave-number region. However, in a system with a FM interaction 
under a static magnetic field, both power laws are confirmed clearly. 

Finally, we comment on the possibility of experimental observation of the $-1$ and $-7/3$ power laws. 
We expect that these power laws can be observed in experiments with $F=1$ $^{87}$Rb (FM interaction) under a static magnetic field if the system size is much larger than $2 \pi / k_{b}$. 
This type of experiment may be feasible because there are currently some experiments with $^{87}$Rb in which magnetic fields are applied and spin density vectors can be observed \cite{Vengalattore}. However, in  experiments with $F=1$ $^{23}$Na (AFM interaction) \cite{Vinit13}, 
it may be difficult to observe these power laws because our numerical calculation does not obtain the clear $-1$ power law in  ST with an AFM interaction.


\begin{thebibliography}{99}
\bibitem{density} If the kinetic energy in an initial state is larger than $Nc_{0} \rho _{0}$, the total density can be disturbed. 
However, in our numerical calculations of Sec. IV, the kinetic energy is smaller than $Nc_{0} \rho _{0}$. Thus, we can neglect the spatial derivative of the total density.
\bibitem{Large_spin} We note the treatment in the case with a large-magnitude  spin vector, which is realized in ST with the FM interaction. In this case, the system is approximately described as a ferromagnetic state, so that the spin vector and the nematic tensor are related through 
$n_{\mu \nu} = (\delta _{\mu \nu} + f_{\mu} f_{\nu})/2$, which eliminates 
the nematic tensor in the spin hydrodynamic equations. Thus, we can perform the Kolmogorov-type dimensional scaling analysis as discussed in \cite{FT12a}, deriving the $-7/3$ power law.
\bibitem{defect} In the one-component BEC, there is a Kelvin wave cascade, in which the Kevin wave transfers the kinetic energy from a low  wave number to a high one. 
This cascade is considered to occur in the wave-number region higher than the wave number corresponding to the mean intervortex distance, 
which is generally larger than the coherence length. Thus, if the motion of spin structures affects the energy transfer in ST, the $-7/3$ power 
law may bend at the wave number corresponding to the mean distance between the spin structures.
\end{thebibliography}
\end{document}